\documentclass[10pt,twocolumn,a4paper]{esaAI}

\title{Magnetogram-to-Magnetogram: Generative Forecasting of Solar Evolution}

\def\authorEmail{francesco.ramunno@fhnw.ch}

\author[1, 2]{Francesco Pio Ramunno\thanks{Corresponding author. E-Mail: \authorEmail}}
\author[3]{Hyun-Jin Jeong}
\author[1]{Stefan Hackstein}
\author[1]{André Csillaghy}
\author[2]{Svyatoslav Voloshynovskiy}
\author[4, 5]{Manolis K. Georgoulis}
\affil[1]{Institute for Data Science, University of Applied Sciences North Western Switzerland (FHNW), 5210 Windisch, Switzerland}
\affil[2]{Department of Computer Science, University of Geneva, 1211 Geneva, Switzerland}
\affil[3]{Department of Astronomy and Space Science, Kyung Hee University, Yongin, 17104, Republic of Korea}
\affil[4]{Johns Hopkins University Applied Physics Laboratory Laurel, MD 20375, USA}
\affil[5]{Research Center for Astronomy and Applied Mathematics of the Academy of Athens, 11527 Athens, Greece}
\newcommand{\customcite}[1]{\citeauthor{#1} (\citeyear{#1})}

\usepackage{siunitx}
\usepackage{tabularx}
\usepackage{makecell} 
\usepackage{soul}
\usepackage{xcolor}

\DeclareUnicodeCharacter{22C6}{\ensuremath{\star}}

\begin{document}

\makeCustomtitle

\begin{abstract}
Investigating the solar magnetic field is crucial to understand the physical processes in the solar interior as well as their effects on the interplanetary environment. We introduce a novel method to predict the evolution of the solar line-of-sight (LoS) magnetogram using image-to-image translation with Denoising Diffusion Probabilistic Models (DDPMs). Our approach combines "computer science metrics" for image quality and "physics metrics" for physical accuracy to evaluate model performance. The results indicate that DDPMs are  effective in maintaining the structural integrity, the dynamic range of solar magnetic fields, the magnetic flux and other physical features such as the size of the active regions, surpassing traditional persistence models, also in flaring situation. We aim to use deep learning not only for visualisation but as an integrative and interactive tool for telescopes, enhancing our understanding of unexpected physical events like solar flares. Future studies will aim to integrate more diverse solar data to refine the accuracy and applicability of our generative model. Visit this \href{https://github.com/fpramunno/MAG2MAG}{https://github.com/fpramunno/MAG2MAG} for the code and \href{https://huggingface.co/spaces/fpramunno/mag2mag}{https://huggingface.co/spaces/fpramunno/mag2mag} for the interactive tool.
\end{abstract}

\section{Introduction}
The Sun's magnetic field is essential for understanding both the solar interior and the interplanetary environment, which includes space weather events close to Earth \cite{leka_a, leka_b, leka_c, Pandey_2023, Yi_et_al2023, GEORGOULIS2024}. This magnetic field originates from within the solar interior and is most accurately measured at heights corresponding to the Sun's visible surface (the photosphere) \cite{Wiegelmann2014}.
\begin{figure*}[t]
    \centering
    \includegraphics[width=2\columnwidth]{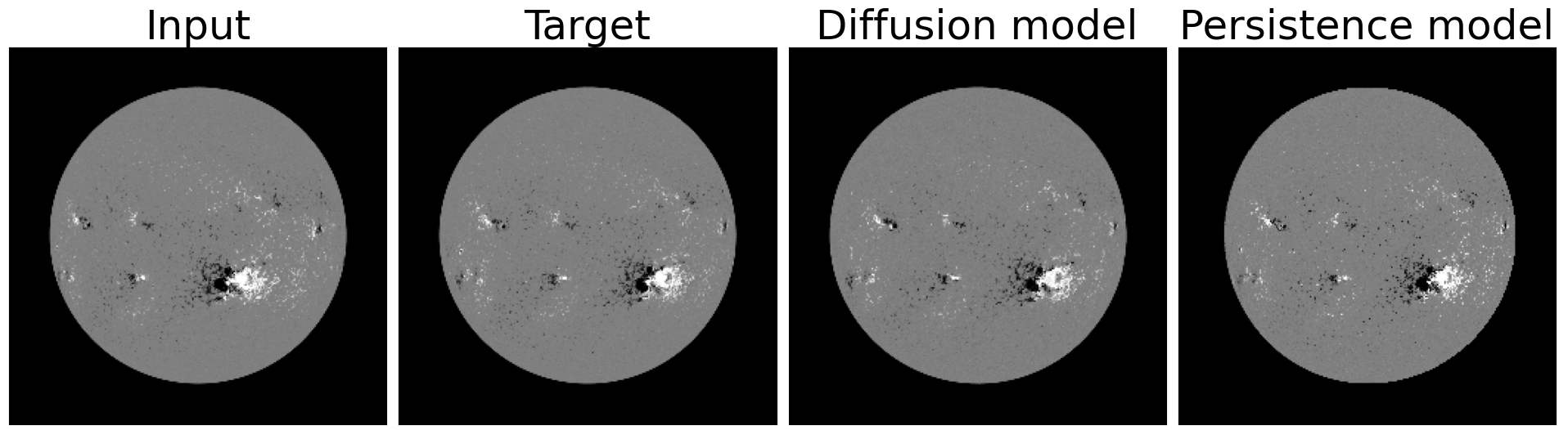}
	\caption{Example of input, target, and a comparison of the model's prediction with the baseline persistence model. Input image from 25 October 2014 at 17:08 UTC.}
	\label{fig:overall_res}
\end{figure*}
Line-of-sight (LoS) magnetograms show the component of the magnetic field vector along the line of sight from the observer to the source and they map the solar photopshere. Solar magnetograms are essential for studying the magnetic field of the Sun and instabilities, including solar flares—rapid, intense bursts in the solar atmosphere that usually last a few minutes but can extend for several hours. Using the high-cadence Helioseismic and Magnetic Imager (HMI) \cite{hmi_instrument} on the Solar Dynamics Observatory (SDO) \cite{sdo_telescope}, it is possible to study solar flare development and prediction for space weather applications \cite{Pandey_2023, son2023three}, as well as the temporal evolution of magnetic surface properties, including magnetic field flux, the size of active regions, their magnetic polarity inversion lines (PILs), etc. \cite{jeong2020solar, jeong2022improved}.

In the last decade, artificial intelligence had a substantial boost in the research field \cite{COLLINS2021102383}, and the most successful machine learning methods in image analysis are built upon deep non-linear neural networks within the field of generative modelling \cite{kingma2022autoencoding, goodfellow2014generative, ho2020denoising}. These networks have been widely applied for various tasks such as image generation \cite{rombach2022highresolution}, super-resolution \cite{saharia2022palette} and image-to-image translation \cite{li2021srdiff}. They have also been applied in the field of heliophysics \cite{ramunno2024solar, Sun_2022, Pandey_2023}.

Image-to-Image translation aims to translate the domain knowledge of an image to another image. In astronomy, it can be used to transfer the knowledge of a telescope to another telescope \cite{s24041151, mariia_drozdova}, allowing us to study objects in another domain, and it can be used to build an "artificial telescope" that analyses an astronomical object from an unseen angle \cite{jeong2022improved, jarolim_2024}.
The study by \customcite{jeong2022improved} demonstrates how STEREO and SDO data can predict solar farside magnetograms as seen by HMI, enhancing its coverage to provide a 360-degree view of the solar photosphere. Thus, image-to-image translation offers promising applications in astronomy if used correctly.

In this work, we use image-to-image translation to predict the full disc LoS magnetogram from HMI using maps from 24 hours before (see Fig \ref{fig:overall_res}). We then evaluate the model's accuracy by measuring physical parameters comparing them with a persistence model that accounts only for the Sun's differential rotation.

This paper is organised as follows: In Sect.~\ref{sec:results} we present our algorithm implementation and its result. In Sect.~\ref{sec:disc_conc} we discuss our results and draw the conclusion.

\section{Implementation and Results}
\label{sec:results}
The three pillars in generative modelling are the Variational Autoencoders (VAEs) \cite{kingma2022autoencoding}, Generative Adversarial Networks (GANs) \cite{goodfellow2014generative} and Denoising Diffusion Probabilistic Models (DDPMs) \cite{ho2020denoising}. The priority criterion for this work is image quality because we intend to test if physics and pixel-level features are respected. For this reason, we do not consider the VAEs despite their speed of inference since they are known to generate blur images. GANs, instead, provide an optimal image quality while being very fast in sampling. However, GANs are notoriously difficult to train and can suffer from limited mode coverage and ambiguous convergence markers. For these reasons, we select the DPPMs due to their high fidelity sample quality and diversity, that can outperform both GANs and VAEs \cite{dhariwal2021diffusion}.
\begin{table*}[h]
\renewcommand{\arraystretch}{1.2}
\begin{center}
    \begin{tabular}{c || c | c | c | c | c | c | c | c | c | c } 
    \hline\hline
    Model & \makecell{PSNR $\uparrow$} & \makecell{SSIM $\uparrow$} & \makecell{LPIPS $\downarrow$} & \makecell{FD TOT\\ Flux} & \makecell{FD NET\\ Flux} & \makecell{AR TOT\\ Flux} & \makecell{AR NET\\ Flux} & \makecell{Size\\ AR} & \makecell{PIL\\ length} & \makecell{Jacc\\ Idx} $\uparrow$ \\
    \hline\hline
    DDPM & $\makecell{\textbf{31.2} \\ \pm \\ \textbf{0.03}}$ & $\makecell{0.7 \\ \pm \\ 0.0004}$ & $\makecell{\textbf{0.03} \\ \pm \\ \textbf{0.0002}}$ & $\makecell{\textbf{0.37} \\ \pm \\ \textbf{0.004}}$ & $\makecell{\textbf{1.25} \\ \pm \\ \textbf{0.0}}$ & $\makecell{\textbf{5.76} \\ \pm \\ \textbf{0.14}}$ & $\makecell{\textbf{13.96} \\ \pm \\ \textbf{11.15}}$ & $\makecell{\textbf{2.9} \\ \pm \\ \textbf{0.51}}$ & $\makecell{\textbf{3.34} \\ \pm \\ \textbf{1.37}}$ & $\makecell{\textbf{0.65} \\ \pm \\ \textbf{0.002}}$ \\
    Persistence & $15$ & $\textbf{0.78}$ & $0.10$ & $2.01$ & $1.25$ & $14.44$ & $42.37$ & $8.09$ & $5.65$ & $0.55$ \\
    \hline\hline
    \end{tabular}
\caption{Results of the experiments based on the metrics explained in Section \ref{sec:results}. The symbol $\uparrow$ indicates that a higher value is preferable, while the symbol $\downarrow$ indicates that a lower value is preferable. The SSIM and the Jaccard Idx are designed such that their maximum value is 1. FD stands for full-disc, TOT means total unsigned magnetic, NET refers to the net value, namely the value with the sign. Except for PSNR, SSIM, LPIPS, and Jaccard index, the metrics measure percentage differences between model predictions and true map values, indicating how far off the predictions are in percentage terms. These metrics have been computed over 10 sets of 250 samples each and the metric value and its associated standard deviation are percentage variations.}
\label{tab:tab_results}
\end{center}
\end{table*}

We use SDO/HMI magnetograms for our purposes \cite{sdo_telescope, hmi_instrument}. HMI records full-disc images of the solar photosphere at \SI{6173}{\angstrom}. In this work, we use LoS magnetograms available in compressed JP2 format from the Helioviewer \cite{helioviewer} website \footnote{\url{https://helioviewer.ias.u-psud.fr/jp2/HMI/}}. The final dataset comprises 43912 paired images from 2013 to 2020 selected with the catalogue by \customcite{Plutino_2023}. The pairs comprise the nearest LoS magnetogram to the flaring peak with a maximum distance of 1 min and its corresponding version 24 h before. Additionally, for computational efficiency, we decreased the image resolution to 256x256.

Our algorithm follows the paper by \customcite{saharia2022palette}. Starting from pure Gaussian noise and the input LoS magnetogram, it tries with 1000 applications of the model to predict how the LoS magnetogram will appear in 24 hours.
\begin{figure}[t]
    \centering
    \includegraphics[width=.50\textwidth]{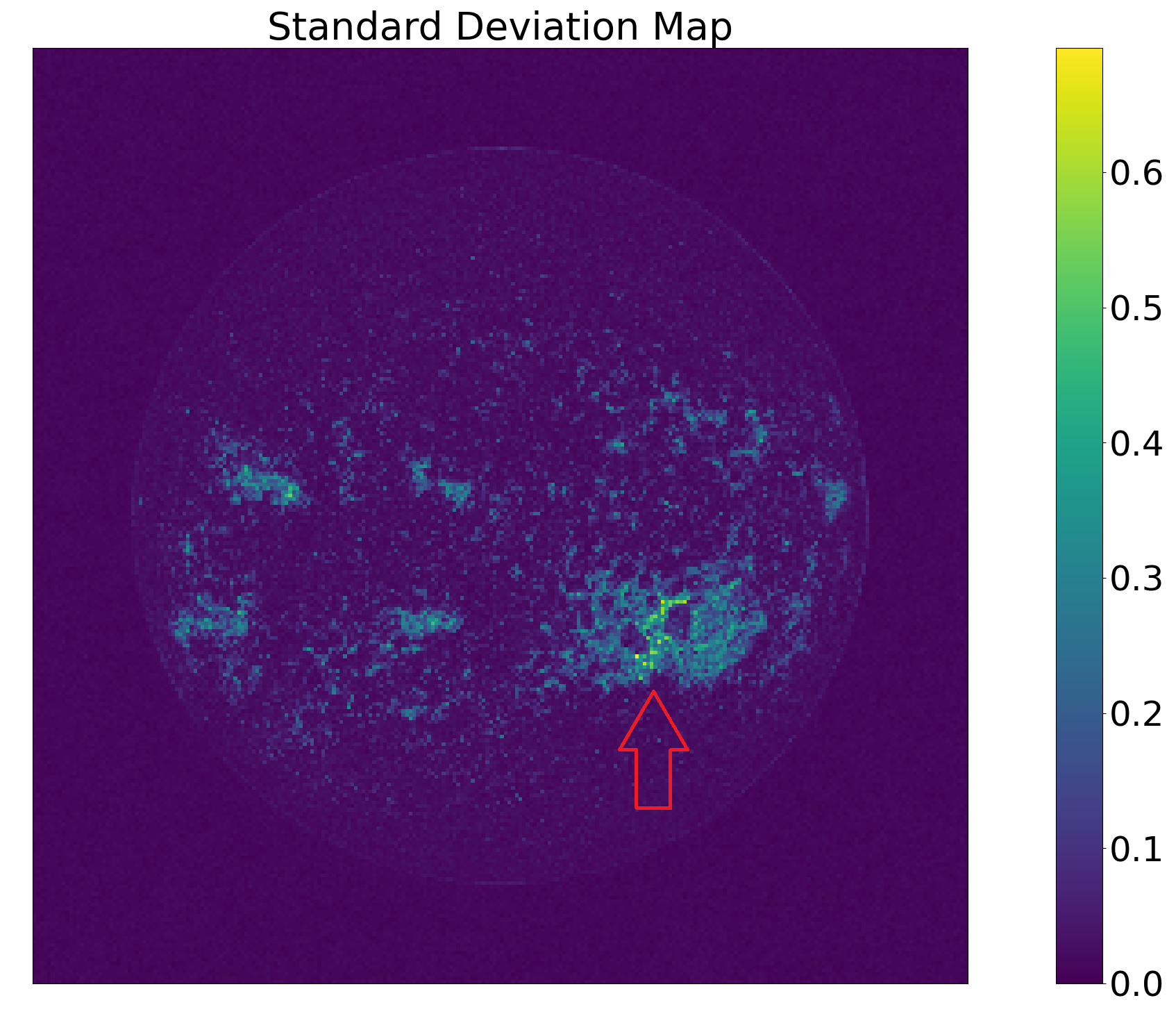}
	\caption{The image shows a standard deviation map derived from 8 repeated (semi-stochastic) model predictions. As expected, the variance shows that the model finds it most difficult to predict the PIL of the AR 12192 (indicated by the arrow), which physically undergoes the most significant changes during a flare. Input image from 24 October 2014 at 17:08 UTC. The standard deviation has been computed on images normalised between -1 and 1, thus the unit of measure is in pixel value.}
	\label{fig:std_image}
\end{figure}
To better evaluate the performance of our model, we employ a set of metrics from the computers science domain that are widely used to assess image quality, such as the Peak-Signal-to-Noise-Ratio (PSNR), the Structural-Similiraty-Index-Measure (SSIM) \cite{ssim} and the Learned Perceptual Image Patch Similarity (LPIPS) distance as a perceptual metric \cite{zhang2018unreasonable}. However, with only those metrics, it is impossible to understand if, while predicting the evolution of the LoS magnetogram, the physics of the processes is also preserved. Thus, we define a "physics metrics" set to judge if the model correctly emulates the physical processes at work. We divide those into the global (full-disc) category and the local (active-region) category. To track all the ARs and compute the metrics, we use the SunPy library \cite{sunpy} to locate all the ARs present at a specific time with their centre through the NOAA (National Oceanic and Atmospheric Administration) AR number. The first physics metrics that we implement are the total magnetic flux for the whole disc and all the ARs, and the net magnetic field flux for the entire disc and all the ARs \cite{Wiegelmann2014} (See section \ref{sec:disc_conc} for the flux calculation from Helioviewer’s JP2 images).
Additionally, we compute the AR size and the PIL length, which are two essential parameters to discriminate if the AR has the potential to flare or not \cite{Sadykov_2017, wang_2020}. To extract the PIL length from the LoS magnetograms, we follow the algorithm proposed in \customcite{pil}. We also use the Jaccard Index \cite{jaccard} to approximate the morphology of ARs. To evaluate the results, we compute the percentage variation (PV) between the metric value obtained on the true LoS magnetogram and the value obtained on the predicted LoS magnetogram.
The results obtained with the DDPM are compared with the persistence model. The persistence model is not a machine learning model, and it considers only the Sun's differential rotation. Hence, the persistence model does not predict the evolution of the AR and their physics features but it carries forward to see how the Sun will look after a given amount of time. We perform this with the SunPy library. 
To evaluate our model, we test on 250 cases randomly selected by the validation set. We apply the model 10 times on the 250 data, obtaining ten sets of 250 predictions of the input data. We compute the metrics for all ten sets and aggregate those results using the mean and the standard deviation. We repeat this analysis because the DDPM is a probabilistic model \cite{ho2020denoising}, thus given the same input we will always obtain slightly different results and so we can retrieve an uncertainty estimation. 
The results are presented in Table \ref{tab:tab_results}.
\section{Discussion}
\label{sec:disc_conc}
As stated in Section \ref{sec:results} we have two sets of metrics: the computer science metrics, which include the PSNR, the SSIM, the LPIPS and the physics metrics. Computer science metrics evaluate the quality of the output regardless of its physical meaning. The PSNR and the SSIM are pixel-based similarity scores. The PSNR measures the ratio of the maximum possible signal power to the power of distorting noise affecting its fidelity. As we can see from Table \ref{tab:tab_results} the DDPM beats the Persistence model, which means that the DDPM is better in predicting the actual dynamic range of the image. The SSIM considers the luminance, contrast, and structure; in this case, the Persistence model is better. However, pixel-level metrics are not always reliable for sample quality \cite{saharia2022palette}, and it has been observed that PSNR and SSIM tend to prefer blurry outputs \cite{menon2020pulse}. Thus, to better evaluate the performance of the model, we use the LPIPS distance, an L2 distance in the latent space of a pre-trained Alexnet model \cite{NIPS2012_c399862d}, and in this case, the DDPM beats the Persistence model. In conclusion, regarding the computer science metrics, the DDPM performs better overall than the Persistence model in terms of dynamic range and overall structure of the predicted image.

Nevertheless, we are interested in more than just the aesthetic of the image: we wish to understand if the model can replicate the nature of some of the physics processes. For this reason, we define a set of metrics that characterise those processes like the magnetic flux, since its changes correlate with the structural complexity of the field, which is fundamental for understanding and predicting events like flares and coronal mass ejections (CMEs) \cite{Wiegelmann2014}. To assess those measurements, we must work on a physics space, which is not the space between 0 and 1 that Helioviewer \cite{helioviewer} shows. To go back into the physics space, we revert the linear transformation used by the Helioviewer project \footnote{https://github.com/Helioviewer-Project/jp2gen/blob/master/idl/sdo/hmi/hvs\_default\_hmi.pro} which goes back to an image whose magnetic field values vary in the range [-250, 250] G.  We compute the total and the net magnetic fluxes, both at the full-disc level and at the ARs level. As shown in Table \ref{tab:tab_results}, the magnetic flux with a full-disc precision has a lower PV than the ARs precision. The magnetic flux is the product of the magnetic field and the area. Thus, the full-disc measure seems more accurate because we are averaging over a more extensive area that contains lower magnetic field values. In this case, both the magnitude and the net magnetic field flux are better estimated by the DDPM, but to truly understand the model capabilities we must go to the AR level. The PV for the AR is higher than the full-disc, as we can see from Table \ref{tab:tab_results}, because in this case we compute the metric over a smaller and more complex area, which is harder to predict. In both cases, the magnitude and the net magnetic field flux, the DDPM is better than the persistence model. Still, there is a difference between the magnitude and the net flux. This difference is shown in the average value and the standard deviation, which determines how uncertain the model is. The magnitude is more straightforward to determine because it represents the overall flux with no interest in the direction of the field.
For this reason, the model encounters more difficulties in its prediction because, given a static input (LoS magnetogram 24 hours before), we are predicting the dynamics of a complex field that changes on a much shorter time scale. In addition to this, the stochasticity of the model has a more significant impact, given that the uncertainty of the prediction is equal to 80 \% of the average predicted value. In future work, we are interested in better exploring the ability to predict dynamic features, including inputs, that take into account different layers of the Sun (e.g. AIA input images), but also to encode features that are correlated with the magnetic field (e.g. velocity fields \cite{Wiegelmann2014}).

To test the limit of the DDPM and demonstrate how it behaves on a finer-grained scale, we use as evaluation parameters also the size of the ARs and PIL length, which widely correlate with events such as flares or CMEs, and the Jaccard index which we use as a proxy for the morphology of the ARs. However, these metrics requires the use of external libraries such as OpenCV \cite{opencv_library} and so additional algorithms, differently from the magnetic flux measurements leading to higher error propagation. To prevent this, we use the same hyper-parameters of the algorithms for all the images, even though to be more precise each image should be characterised by its parameters. Nonetheless, we see from Table \ref{tab:tab_results} that for all of these metrics, the DDPM prediction is better than the Persistence model.
Consequently, the output given by the DDPM can be used not only as a visualisation technique but also as a scientific result that can be studied and analysed. Moreover, the training and validation datasets consist of image pairs captured near the flare peak and 24 hours earlier. The model's accurate predictions of these images demonstrate its ability to forecast the Sun's behaviour during flare events. Since the length of the PIL and the AR area are reliable indicators of potential flares, accurately predicting these values suggests that the model holds essential information about flare events 24 hours in advance. Nevertheless, flare prediction is not the topic of this project but in future work we aim on using explainable AI to understand the model behaviour \cite{panos} during these events. Moreover, due to the probabilistic nature of the DDPM, it is interesting to check what is the most uncertain part of the LoS magnetogram to predict. We input the image 24 hours before the X1.0 flare happened on 25 October 2014 from the AR 12192, and we predict it eight times. To visualise the uncertainty of the model, we concatenate the eight predictions along the channel dimension and then compute the standard deviation per pixel. This process outputs an image where the higher the value is, the bigger the standard deviation. Thus, the model is more uncertain on that pixel prediction. As shown in Figure \ref{fig:std_image}, the values are higher in correspondence with the AR 12192, which is the flaring region, and most importantly, it is more uncertain in the region of polarity inversion (RoPI). This result is consistent with our physical knowledge as shown by \customcite{kleint_lucia} that photospheric changes during an event like a flare are predominantly located near the PIL. Considering the computational constraints (e.g. size reduction to 256x256 pixels) and the initial design of the model (e.g. using only one LoS magnetogram), the performance on the physics metrics of the DDPM is encouraging. Thus, we are interested in expanding this in future work to test if injecting different types of information (e.g. velocity fields, AIA channels, LoS magnetograms of multiple days) helps the model to understand the evolution of these dynamic processes better and thus explore which processes cause the magnetic field in the lower solar atmosphere to change, which is not fully understood yet. 

\begin{acknowledgements}
      This research was partially funded by the SNF Sinergia project (CRSII5-193716): Robust Deep Density Models for High-Energy Particle Physics and Solar Flare Analysis (RODEM).
\end{acknowledgements}

\printbibliography
\end{document}